\documentclass[aps,prb,10pt,twocolumn,superscriptaddress,floatfix,amsmath,amssymb]{revtex4-1}

\usepackage{graphicx,xcolor}
\usepackage{mathtools}
\usepackage{hyperref}
\usepackage{epstopdf}
\usepackage[section]{placeins}

\begin{document}

\title{Charge ordering and non-local correlations in the doped extended Hubbard model}

\author{Hanna Terletska} 
\affiliation{%
 Department of Physics, Middle Tennessee State University, Murfreesboro, TN 37132, USA
}%
\affiliation{Department of Physics, University of Michigan, Ann Arbor, MI 48109, USA} 
\author{Tianran Chen} 
\affiliation{Department of Physics, West Chester University of Pennsylvania, West Chester, PA 19383, USA}
\author{Joseph Paki} 
\affiliation{Department of Physics, University of Michigan, Ann Arbor, MI 48109, USA}
\author{Emanuel Gull}
\affiliation{Department of Physics, University of Michigan, Ann Arbor, MI 48109, USA}

\date{\today}

\begin{abstract}
We study the extended Hubbard model away from half-filling on a two-dimensional square lattice using cluster dynamical mean field theory on clusters of size $8$. We show that the model exhibits metallic, compressible charge ordered, and insulating charge ordered phases. We determine the location of the charge ordering phase transition line at  finite temperature and the properties of the phases as a function of doping, temperature, local interaction, and nearest neighbor interaction. An analysis of the energetics of the charge order transition shows that the charge ordering transition mainly results in a rearrangement of local and non-local potential energy. We show the doping evolution of the spectral function from the isotropic metal via a charge ordered metal to a charge ordered insulator with a big gap, and study finite size effects of the approximation.
\end{abstract}

\maketitle

\section{Introduction} 
The energetic competition between electron repulsion due to the Coulomb interaction, which tends to localize electrons, and kinetic effects, which favor electrons itineracy, leads to a rich interplay of competing phases in strongly correlated systems, where both contributions are of comparable magnitude.\cite{Imada}

In lattice model calculations, the Coulomb interaction is often approximated as a purely local term, resulting in models such as the Hubbard model.\cite{Hubbard63,LeBlanc15,LeBlanc13} However, non-local terms of the interaction are always present in real systems, necessitating a careful treatment of {\it all} interactions for quantitative results,\cite{Motta17,Lan17} along with the development of powerful numerical methods that treat both general interactions and strong correlations.\cite{Lan17B,Zgid17,Kananenka15} As these non-local interactions increase in strength, they lead to qualitatively new physics, including symmetry-broken charge-ordered states.

Charge ordered states are ubiquitous in nature. Since their early observation by \textcite{Verwey39} in magnetites, they have been found in  Wigner crystals,\cite{Wigner,Lenac} high $T_c$ cuprates, \cite{Tranquda,Davis,Yazdani,SilvaNeto,Lena} manganites,\cite{Tokura1995,Cheong1996,Cheong2002,Dagotto2001} cobaltates,\cite{cobaltates} nickelates,\cite{NiO2015,Ni_Cheong,Ni_Zhang,LaNiO} two-dimensional organic materials, \cite{Jerome,Kanoda-organic-Wigner,Hotta,Dressel} in La$_{1-x}$Sr$_x$FeO$_3$,\cite{LSFO1981,LSFO2007} layered dichalcogenides,\cite{dichalcogenides} and other, including quasi-one-dimensional\cite{1d_1,1d_2} systems. 

Charge order effects resulting from electronic interactions can be studied theoretically on model systems that are both simple enough that different physical phenomena can be disentangled, and complex enough that they exhibit the salient aspects of correlation physics in the presence of non-local interactions. The extended Hubbard model, which includes nearest neighbor density-density interactions in addition to the local Coulomb repulsion, but which neglects all non-density-density terms and all Coulomb interactions beyond the nearest neighbor, is such a minimal model.

In a previous paper, Ref.~\onlinecite{Terletska17}, we performed an analysis of the two-dimensional extended Hubbard model within the dynamical cluster approximation (DCA)\cite{Maier05} on clusters of size $8$ and larger. We performed a systematic study of the properties of the charge ordered and charge disordered uniform phase at half-filling and for finite temperature. Our results showed that the increase of intersite interactions $V$ for fixed local interactions $U$ leads to the establishment of a charge-ordered (CO) phase which is characterized by a checkerboard arrangement of electrons with nonzero staggered density. The charge ordered phase persists up to a critical temperature $T_{CO}$ that depends strongly on the strength of $V$ and $U$, and can be destroyed by increasing $U$ for fixed $V$.

In this paper, we show how these findings change when the number of particles is doped away from half-filling. We explore the phase diagram as doping, non-local interaction $V$, local interaction $U$, and temperature $T$ are varied. We analyze the behavior of spectral functions and the gap formation mechanism, the behavior of the order parameter, compressibility effects, as well as the energetic competition between kinetic, local potential, and non-local potential terms in detail. Our analysis shows that the establishment of a charge ordered phase rapidly lowers the non-local interaction energy, at the cost of substantially raising the local interaction contribution. While the results analyze the competition between charge order and the isotropic metallic state in detail, our interaction strengths stay below the Mott insulating limit and long-range antiferromagnetism is suppressed.

The remainder of this paper will proceed as follows. In Sec.~\ref{sec:ModelMethod}, we will give an overview of the model and  our method. Sec.~\ref{sec:PD} will analyze the phase diagram in detail. Sec.~\ref{sec:OP} will discuss the evolution of the order parameter, Sec.~\ref{sec:SF} the spectral function, and Sec.~\ref{sec:Energy} show the energetics of the charge order phase transition as a function of doping. Section \ref{sec:FS} contains a brief analysis of finite size effects, and Section \ref{sec:Conc} will present conclusions.

\section{Model and method} \label{sec:ModelMethod}
In this paper, we use the model, method, and formalism of Ref.~\onlinecite{Terletska17}. We repeat some of the method aspects here in order to keep it self-contained but refer the reader to Ref.~\onlinecite{Terletska17} for a more detailed explanation.

The extended Hubbard model on a two-dimensional square lattice is given by the Hamiltonian
\begin{align}
H&=-t\sum_{\langle ij\rangle,\sigma}\left ( c_{i\sigma}^{\dagger}c_{j\sigma}+c_{j\sigma}^{\dagger}c_{i\sigma}\right )+U\sum_{i}n_{i\uparrow}n_{i\downarrow}  \nonumber \\
&+\frac{V}{2}\sum_{\langle ij\rangle,\sigma\sigma'}n_{i\sigma}n_{j\sigma'}-\tilde{\mu}\sum_{i\sigma}n_{i\sigma},
\label{Hamiltonian}
\end{align}
where $t$ is the nearest-neighbor hopping amplitude, $U$ and $V$ are the on-site and nearest neighbor Coulomb interactions, respectively, and $\tilde{\mu}$ denotes the chemical potential. $c_{i\sigma}^{\dagger} (c_{i\sigma})$ is the creation (annihilation) operator for a particle with spin $\sigma$ on lattice site $i$, and $n_{i\sigma}=c_{i\sigma}^{\dagger}c_{i\sigma}$ is the number operator on site $i$. The system is half filled at $\tilde{\mu}=\mu_\text{HF}=\frac{U}{2}+zV$, ($z$ is the coordination number), and in the remainder of this paper we specify the chemical potential with respect to half filling, such that $\tilde\mu=\mu_\text{HF}+\mu$ and the system is half-filled for $\mu=0$. Throughout this paper we use dimensionless units $U/t$, $V/t$, $\beta t$, and $\mu/t$, and set $t=1$.

Early studies of the extended Hubbard model in two dimensions with lattice Monte Carlo,\cite{Zhang89} exact diagonalization,\cite{Callaway90,Ohta94} weak\cite{Dongen94Weak} and strong\cite{Dongen94Strong} coupling perturbation theory as well as high temperature series expansion\cite{Bartkowiak95} mainly focused on the interplay of spin, charge, and superconducting degrees of freedom.
Later calculations, some of them performed with non-perturbative embedding methods, were primarily motivated by four aspects: by applications to the physics of the organic superconductors,\cite{Merino04,Merino07} aspects of which are believed to be described by a quarter filled extended Hubbard model; by
the exploration of superconducting properties in the presence of non-local interactions;\cite{Onari04,Davoudi06,Davoudi07,Davoudi08,Husemann12,Senechal13,Huang13,Plonka15,Reymbaut16,Jiang17}  by methodological development\cite{Bolech03,Aichhorn04,Loon14,Loon15}; 
and by the fundamental question of the `screening' effect that non-local interactions have on the normal state physics of models with large local interactions.\cite{Ayral13,Loon14,Loon15,Loon16,Sawatzky} A  study of the charge order phase away from half-filling on the two-dimensional extended Hubbard model over a large doping regime was done in Ref.~\onlinecite{kapcia} using the local extended dynamical mean field approximation and was limited to zero temperature. Very recently, a study of the model in the very large doping regime on a triangular lattice has appeared, making connection to Na$_{x}$CoO$_2$.\cite{Chauvin17}

We approximate the solution of this lattice model with the Dynamical Cluster Approximation (DCA).\cite{Hettler98,Maier05} This method captures short-ranged spatial correlations non-perturbatively, while all correlations outside the cluster are neglected, and can enter the symmetry broken state.\cite{Maier05,Fuchs11} The method is controlled, in the sense that the inverse $1/N_c$ of the cluster size $N_c$ is a small parameter, and becomes exact in the limit of $N_c \rightarrow \infty$. Results obtained within the dynamical cluster approximation on the Hubbard model with only local interactions are now regularly extrapolated to the thermodynamic limit,\cite{Maier05B, Kent05,Gull11,Fuchs11,Kozik10,LeBlanc13,LeBlanc15} where they provide unbiased solutions of interacting fermionic lattice models that have been validated against other numerical methods.\cite{LeBlanc15} 
They are also used as reference data to calibrate and cross-validate ultracold atomic gas experiments.\cite{Imriska14} However, in this paper we mainly present results for $N_c=8$, which in the two-dimensional Hubbard model has been found to capture much of the interesting momentum-dependent pseudogap physics.\cite{Werner09,Gull09,Gull10} Sec.~\ref{sec:FS} contains a brief discussion of finite size effects beyond $N_c=8$.

The DCA is based on a partitioning of the Brillouin zone into $N_c$ patches each centered around a momentum $K$.\cite{Maier05} The many-body self-energy $\Sigma(k, \omega)$ is then expanded into basis functions $\phi_K(k)$, $K=1, \dots, N_c$ which are chosen to be  $1$ for $k$ inside `patch' $K$ and zero otherwise, so that the self-energy is approximated as $\Sigma(k,\omega) \approx \sum_K^{N_c} \phi_K(k) \Sigma(K,\omega)$.\cite{Fuhrmann07} Self-energies of this form can then be obtained from the self-consistent solution of a cluster quantum impurity problem.

To explicitly study the effect of charge ordering we extend our Hamiltonian with a symmetry breaking term by adding a staggered chemical potential $\mu_i=\mu_0 e^{iQr_i}$ with $Q=(\pi,\pi)$ to Eq. (\ref{Hamiltonian}):
\begin{equation}
H_{\mu_0}=H+\sum_{i\sigma} \mu_i n_{i\sigma}
\label{Eq.2}
\end{equation}
This term breaks the original bipartite lattice into two sub-lattices $A$ and $B$ with $\mu_i=\pm \mu_0$ for sub-lattice $A(B)$ respectively, thereby doubling the unit cell. In this paper, we only consider solutions for $\mu_0 \rightarrow 0$.

The main numerical work in solving the DCA equations consists of solving the quantum cluster problem, {\it i.e.} obtaining an approximate self-energy $\Sigma(K,i\omega_n)$ for a given non-interacting Green's function.
We use the continuous time auxiliary field quantum Monte Carlo algorithm (CTAUX) as a cluster solver.\cite{Gull08} CTAUX is based on the combination of an interaction expansion\cite{Rubtsov05} combined with an auxiliary field decomposition of the interaction vertices. A detailed description of the algorithm is given in Ref.~\onlinecite{Gull11,Gull11_RMP} and the adaptation to non-local density-density interactions is described in Ref.~\onlinecite{Terletska17}. Unlike most other methods, CT-AUX treats non-local terms to all orders in a non-perturbative fashion. An explicit frequency dependence of `effective' `screened' interactions, as required in methods based on the single site dynamical mean field theory,\cite{edmft1,edmft2,edmft3,Loon14,Ayral13} does not arise.

\section{Phase diagram} \label{sec:PD}
We first briefly discuss the results obtained at half filling $(\mu=0)$ and presented in Ref.~\onlinecite{Terletska17}. The phase diagram in the space of on-site interaction $U$ and nearest neighbor interaction $V$ shows metallic behavior for small $U$ and small $V$, Mott insulating behavior for large $U$ and small $V$, and charge order for large $V$. This basic shape of the phase diagram is consistent within a large range of methods, and in particular with recent results using extended dynamical mean field theory,\cite{Medvedeva17} the $GW$ approximation in combination with dynamical mean field theory,\cite{Ayral13,Ayral17} and the so-called dual boson perturbation theory.\cite{Loon14} In contrast to the predictions from early analytic theories,\cite{Bari71,Wolff83,Yan93} a non-zero strength of $V$ is required to drive the system to the ordered phase at $U=0$. Upon increasing the interaction strength, the charge order line stays above the mean field prediction of $U=4V$ but, at least for $U/t$ up to $1$, closely approaches it.

\begin{figure}[tbh]
\includegraphics[width=0.85\columnwidth]{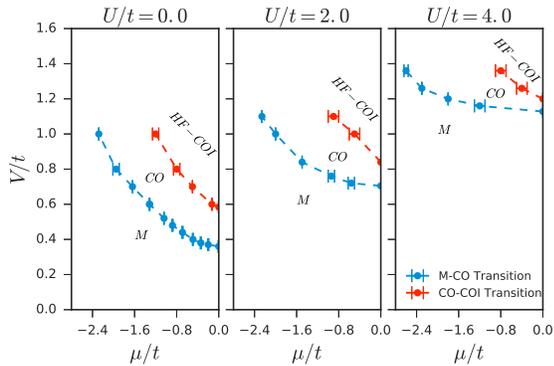}
\caption{Phase diagram showing half-filled charge ordered insulator (HF-COI), charge-ordered metal (CO), and isotropic metal (M), in the space of nearest neighbor interaction $V$ and chemical potential $\mu$. Left panel: on-site interaction $U/t=0$. Middle panel: $U/t=2.0$. Right panel: $U/t=4$. All data are obtained for cluster size $N_c=8$ and temperature $T/t=0.32$.
\label{fig:PDVvsMu_new}
}
\end{figure} 

We now turn to the doping evolution of the phase diagram. Fig.~\ref{fig:PDVvsMu_new} shows the evolution of this charge order phase boundary upon varying the chemical potential $\mu$, where $\mu=0$ denotes the half-filled state. While the model is particle-hole symmetric around $\mu=0$, we focus in this paper on hole doping and denote $x=1-n$ as doping, where $n$ denotes the density. Results for the system without on-site interaction ($U=0$) are shown in the left panel. Weak ($U/t=2$) and intermediate ($U/t=4$) interaction strength results are shown in the middle and right panel. All results are obtained at a temperature of $T/t=0.32$.

\begin{figure}[tbh]
\includegraphics[width=0.85\columnwidth]{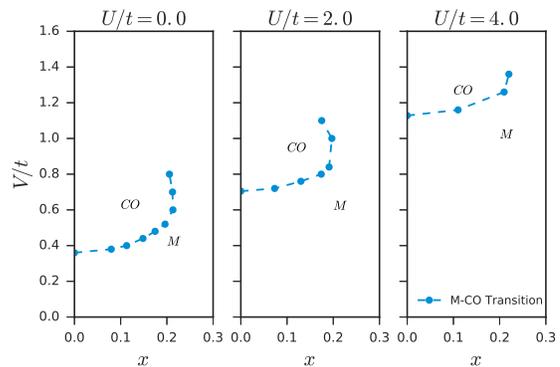}
\caption{Phase diagram in the space of nearest neighbor interaction $V$ and doping $x=1-n$ showing charge-ordered metal (CO), and isotropic metal (M) phases.  Left panel: on-site interaction $U/t=0$. Middle panel: $U/t=2.0$. Right panel: $U/t=4$. All data are obtained for cluster size $N_c=8$ and temperature $T/t=0.32$.
}
\label{fig:PDVvsN_new}
\end{figure}

Consistent with Ref.~\onlinecite{Terletska17} and earlier results,\cite{Loon14,Loon15,Ayral13} a non-zero interaction strength of $V/t \sim 0.4$ is needed to establish charge order. As temperatures are higher than in Ref.~\onlinecite{Terletska17}, the minimal  interaction strength for the onset of charge-order  is larger. As the chemical potential is increased, the charge-order phase boundary shifts to larger $V$  (blue line).  Raising the on-site interaction to weak ($U/t=2$) and moderate ($U/t=4$) strength shifts the onset of charge order gradually to higher $V$.

At high interaction strength $V$, the charge-ordered state is an incompressible, half-filled band-like insulator with a large gap in the density of state. Fig.~\ref{fig:PDVvsMu_new} denotes this regime as HF-COI (Half Filled Charge Ordered Insulator). This regime is separated from the uniform (isotropic) metallic phase by a compressible `metallic' charge ordered state, which we denote as CO (red line).

At the temperatures studied, the transition from non-half-filled to half-filled state and the transition from charge-ordered to isotropic state are all continuous. No jump in order parameter or hysteresis could be identified. The model is known to exhibit first order transitions between the metallic state and the isotropic Mott insulator, both in single-site approximations to the extended Hubbard model\cite{Schuler17} and in cluster approximations to the Hubbard model without non-local interactions.\cite{Gull08,Gull09} However, these phase transitions take place at local interaction strengths that are larger than the ones studied here.

Fig.~\ref{fig:PDVvsN_new} shows the data of Fig.~\ref{fig:PDVvsMu_new} replotted against doping $x=1-n$ rather than chemical potential. The left panel shows $U/t=0$, the middle panel  $U/t=2$, and the right panel $U/t=4$. In this representation, the half-filled charge order insulating regime is compressed to the $x=0$ line. It is evident that once the critical $V$ for charge order is reached, a charge ordered phase is established almost independently of doping, as long as $x\lesssim 20\%$. Further increase of doping eventually leads to the destruction of the charge ordered phase, and the system becomes a uniform metal. Upon increasing the local interaction strength $U$, the critical value of nearest neighbor interaction $V$ increases due to the competition between local and non-local interactions, but the phase boundary remains near $x \sim 20\%$  and a slight back-bending is visible for larger $U$, indicating that the regime of doping shrinks as $V$ is further increased. Numerical difficulties with the impurity solver currently make regimes of larger non-local interaction strength inaccessible.

\begin{figure}[tbh]
\includegraphics[width=0.85\columnwidth]{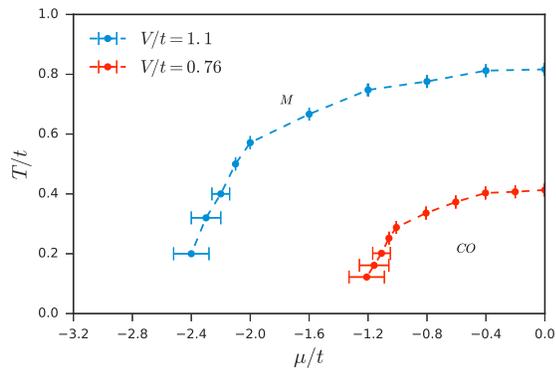}
\caption{Phase boundaries for $V/t=1.1$ (blue line) and $V/t=0.76$ (red line) in the space of temperature $T$ and chemical potential $\mu$ showing charge-ordered (CO) and isotropic metal (M) phases.  All data are obtained for cluster size $N_c=8$ and local interaction strength $U/t=2.0$.
 }
\label{fig:TvMu_PD}
\end{figure}

\begin{figure}[bt]
\includegraphics[width=0.85\columnwidth]{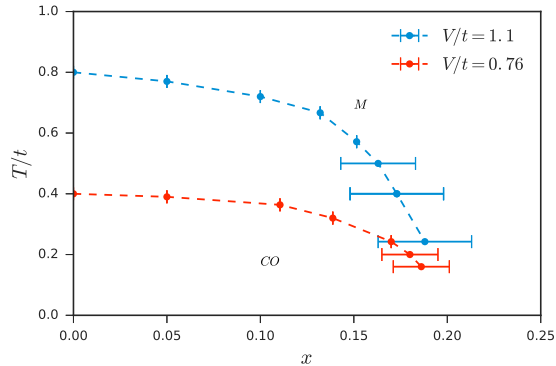}
\caption{Phase boundaries for $V/t=1.1$ (blue line) and $V/t=0.76$ (red line) in the space of temperature $T$ and doping $x$ showing charge-ordered (CO) and isotropic metal (M) phases, for $U/t=2.0$.  All data are obtained for cluster size $N_c=8$. 
\label{fig:Tvx_PD}}
\end{figure}

In order to establish the temperature and doping parameter space for the charge order phase, which corresponds to the phase diagram typically measured in experiments,\cite{morosan} we now explore the temperature dependence of the charge order regime shown in Figs.~\ref{fig:TvMu_PD} and \ref{fig:Tvx_PD}, as a function of chemical potential $\mu$ and doping $x$. The data are obtained for two representative nearest-neighbor interaction strengths $V/t=0.76$ and $V/t=1.1$, which correspond to CO metal and HF-COI, respectively. At high temperature, the system is in a metallic state ($M$), at low temperature in a charge ordered state (CO). For doping up to $10\%$, the charge order onset temperature is almost independent of doping. As doping is gradually increased beyond $15\%$, the onset temperature is rapidly suppressed and, within the parameter range we could reliably study, no charge order is found beyond around $20\%$ doping.

\begin{figure}[bth]
\includegraphics[width=0.85\columnwidth]
{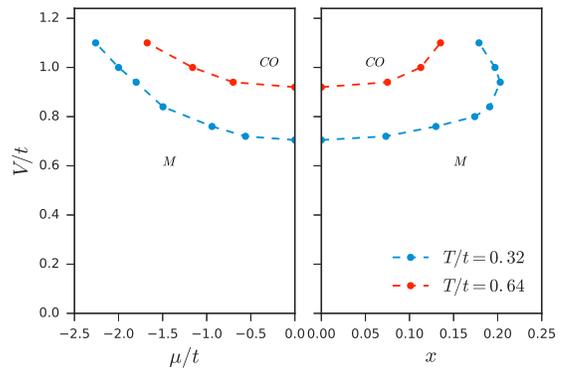}
\caption{Phase boundaries for $T/t=0.64$ (blue line) and $T/t=0.32$ (red line) in the space of nearest-neighbor interaction $V$ and chemical potential $\mu$ (left panel) / doping $x$ (right panel) showing charge-ordered (CO) and isotropic metal (M) phases.  All data are obtained for clusters of size $N_c=8$ and local interaction strength $U/t=2$.
\label{fig:Vvmu_PD}
}
\end{figure}

In analogy to the data shown in Figs.~\ref{fig:PDVvsMu_new} and \ref{fig:PDVvsN_new}, the data in Fig.~\ref{fig:Vvmu_PD} shows the phase boundary between charge order and metallic states as a function of non-local interaction $V$ and chemical potential $\mu$. To highlight the temperature dependence, we also show data at a temperature that is twice as large. As in Fig.~\ref{fig:TvMu_PD}, the regime supporting charge order shrinks substantially as temperature is raised and thermal fluctuations suppress charge order. In addition, the `reentrant' backbending behavior as a function of $V$ is only visible at low temperature.

\section{Order Parameter}\label{sec:OP}
\begin{figure}[tb]
\includegraphics[width=0.85\columnwidth]{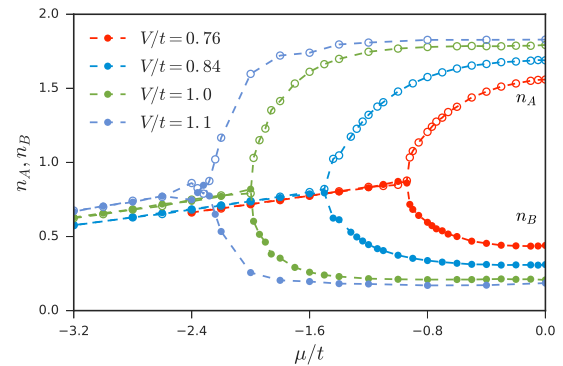}
\caption{Sublattice densities $n_A$ and $n_B$ as a function of chemical potential $\mu$, showing spontaneous establishment of charge-order symmetry breaking, for $U/t=2.0$, $Nc=8$, temperature $T/t=0.32$, and nearest-neighbor interaction strengths $V$ indicated.}
\label{fig:subdens_vs_mu}
\end{figure}

The phase boundaries of Figs.~\ref{fig:PDVvsMu_new} through \ref{fig:Vvmu_PD} were obtained 
by analyzing the site-dependent density as a function of external parameters such as $\mu$ or $V$. A typical example is given in Fig.~\ref{fig:subdens_vs_mu}, where the densities in sublattice $A$ and $B$, $n_A$ and $n_B$, are plotted as a function of chemical potential, for on-site interaction strength $U/t=2$ and four different non-local interaction strengths. As the chemical potential is raised towards half filling, a spontaneous symmetry breaking of the sublattice densities is visible, indicating the establishment of charge order. As mentioned previously, no first-order hysteresis could be found in our simulations, indicating that all transitions are continuous. Larger nearest-neighbor interactions lead to an earlier onset of the charge order and correspondingly a larger polarization of the density.

\begin{figure}[bt]
\includegraphics[width=0.85\columnwidth]{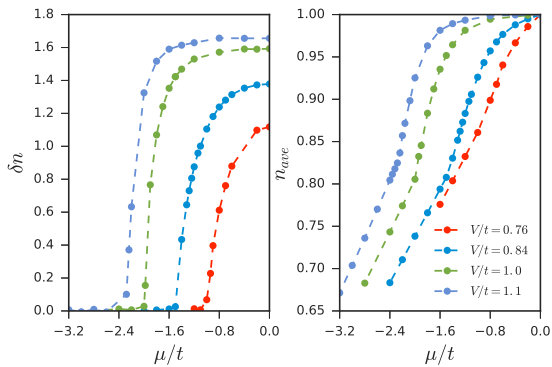}
\caption{(Left) Order parameter $\delta n=n_A-n_B$ and (Right) the average total density $n_{ave}=\frac{n_A+n_B}{2}$ as function of chemical potential $\mu$ for $V/t=0.76, 0.84, 1,$ and $1.1$, at temperature $T/t=0.32$, $U/t=2$, and on clusters of size $N_c=8$.
\label{fig:order_parameter_totdens}
}
\end{figure} 

The  order parameter for charge order, $\delta n=n_A-n_B$, is shown in the left panel of Fig.~\ref{fig:order_parameter_totdens}, and the total density $n_\text{ave}=\frac{1}{2}\left(n_A+n_B\right)$ in the right panel of Fig.~\ref{fig:order_parameter_totdens}. In the uniform phase, $\delta n=0$, whereas $\delta n \neq 0$ in the charge order phase. The data are shown as a function of chemical potential $\mu$ and for a range of non-local interactions $V$, at constant on-site interaction $U/t=2$ and temperature $T/t=0.32$. These data are obtained directly from the sublattice densities shown in Fig.~\ref{fig:subdens_vs_mu}.

The total density $n_{ave}$, shown in the right panel of Fig.~\ref{fig:order_parameter_totdens}, shows a clear deviation from the linear slope at the position where charge order is established ($\delta n \neq 0$). In addition, for larger  inter-site interaction strength, $V/t=0.76$ and $V/t=0.84$, a pinning of the $n_\text{ave}$ vs $\mu$ curve to half filling with $n_{ave}=1$ is visible, indicating an incompressible band insulator-like state (with a robust gap in the density of states) near half filling. As the non-local interaction strength $V$ is increased, the slope of the $n_{ave}(\mu)$ curve rapidly increases on the the ordered side in the vicinity of the phase transition, indicating that first-order coexistence between CO and uniform metallic state may be possible at even larger $V$. 

The HF-COI phase boundary of Fig.~\ref{fig:PDVvsMu_new} (red line) was determined by setting a cutoff value of $n \geq 0.995$. As we will show in Sec.~\ref{sec:SF}, this criterion based on the density also coincides with the region in which the system has a large insulating gap.

The increase of the slope of the $n_{ave}(\mu)$ curve in Fig.~\ref{fig:order_parameter_totdens} and, consequently, the narrowing of the region between the charge ordered insulator ($n_{ave}=1$) and isotropic metal ($\delta n=0$) is directly responsible for the reentrant behavior observed in Figs.~\ref{fig:PDVvsN_new} and \ref{fig:Tvx_PD}.

\begin{figure}[htb]
\includegraphics[width=0.85\columnwidth]{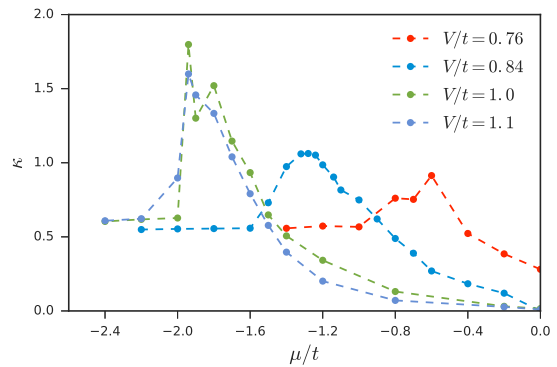}
\caption{Compressibility $\kappa=\frac{\partial n_{ave}}{\partial \mu}$ as a function of chemical potential $\mu$ for nearest neighbor strengths indicated. Data obtained for $U/t=2.0$ on a cluster with $N_c=8$ and at temperature $T/t=0.32$.
}
\label{fig:compressibility}
\end{figure} 

The compressibility $\kappa=\frac{\partial n_{ave}}{\partial \mu}$, Fig.~\ref{fig:compressibility}, here obtained via numerical derivative of the $n_{ave}(\mu)$ curve, shows that the compressibility exhibits a clear maximum at the charge order onset. This maximum becomes more pronounced as $V$ is increased. The value of $\kappa$ quickly approaches a roughly constant large-$\mu$ value on the uniform side of the transition.
Consistent with expectation, the half-filled charge ordered insulating state shows a strongly suppressed compressibility. 

\begin{figure}[tbh]
\includegraphics[width=0.85\columnwidth]
{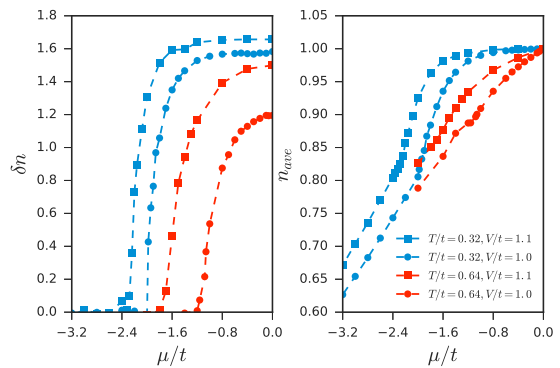}
\caption{Order parameter $\delta n$ (left panel) and total average density $n_{av}$ (right panel) at temperature $T/t=0.32$ (blue curves) and $T/t=0.64$ (red curves) for $U/t=2$, $N_c=8$, and for interactions $V/t=1.1$ (squares) and $V/t=1.0$ (circles) respectively.
}
\label{fig:delta_n_nav_new}
\end{figure}
The temperature dependence of the order parameter $\delta n$ is shown in Fig.~\ref{fig:delta_n_nav_new}. Similar to increasing $V$, decreasing temperature increases the size of the order parameter and the region supporting charge order phase. The half-filled charge ordered insulator is only present in our data for $T/t=0.32$, showing that high temperatures melt the half-filled insulator before the charge order is fully destroyed.

\section{Spectral functions}\label{sec:SF}
\begin{figure*}[t!]
\includegraphics[width=\textwidth]{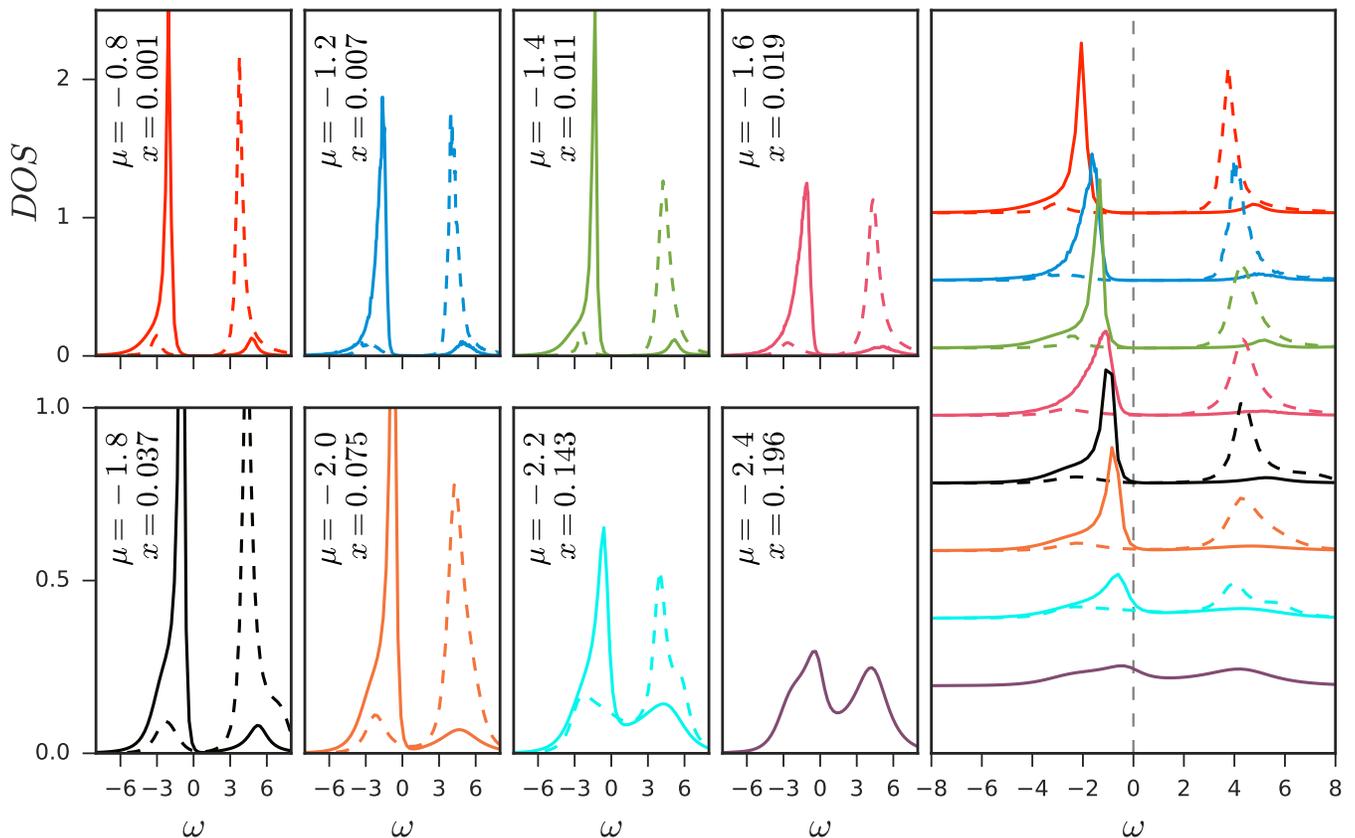}
\caption{Density of states (DOS) for a set of chemical potentials from $\mu/t=-0.8$ to $\mu/t=-2.4$, as a function of frequency. Full lines: sublattice $A$. Dashed lines: sublattice $B$. Values are obtained for $T/t=0.32$, $V/t=1.1$, and $U/t=2.0$. For corresponding densities see Fig.~\ref{fig:order_parameter_totdens}, for compressibilities see Fig.~\ref{fig:compressibility}.
}
\label{fig:dos}
\end{figure*}

Analytically continued local spectral functions, Fig.~\ref{fig:dos}, give further insight into the evolution of the charge order transition as a function of doping. For clarity, we limit ourselves to a single scan in doping, using fixed values of $U/t=2.0$, $V/t=1.1$, and $T/t=0.32$. This is the data corresponding to Fig.\ref{fig:subdens_vs_mu}.

At large doping (not shown), the system is an isotropic Fermi liquid with a small self-energy near zero, so that no suppression of the density of states near zero is visible. As $x$ is lowered from the uniform side towards the onset of charge order (purple panel, $x=0.196$), the density of states develops a clear suppression near zero indicative of strong charge order fluctuations. However, a symmetry breaking is not yet visible. Reduction of $x$ to $0.143$ (light blue panel) and $0.075$ (orange panel) shows the establishment of symmetry breaking but a finite density of states remains at the Fermi energy, indicating a charge order metal. This region coincides with the region of large compressibility visible in Fig.~\ref{fig:compressibility}. As doping is further reduced, the peak-to-peak distance of the minority and majority occupancy spectral functions (full and dashed line) gradually widens and a full gap is established by $x=0.037$. At this point, further doping transfers spectral weight from below the gap to above the gap, while the lower gap edge stays pinned to the Fermi energy and a large density of states is present just below of the Fermi energy. Finally, as $x$ reaches values near zero, the Fermi energy detaches from the gap edge and moves towards the middle of the gap (at $x=0$, not shown), while the minority and majority bands become fully particle-hole symmetric.

The results in Fig.~\ref{fig:dos} are obtained via analytic continuation from noisy quantum Monte Carlo results. In this instance, we used a Pad\'{e} continuation method, which we crosschecked against an implementation of the stochastic analytic continuation method. While analytic continuation does not capture subtle features of the spectral functions, it is generally reliable for the global features (existence of a gap or of the first major peak, weight integrated over a large area, etc) interpreted in the paragraph above.

\begin{figure*}[bth]
\includegraphics[width=0.85\textwidth]{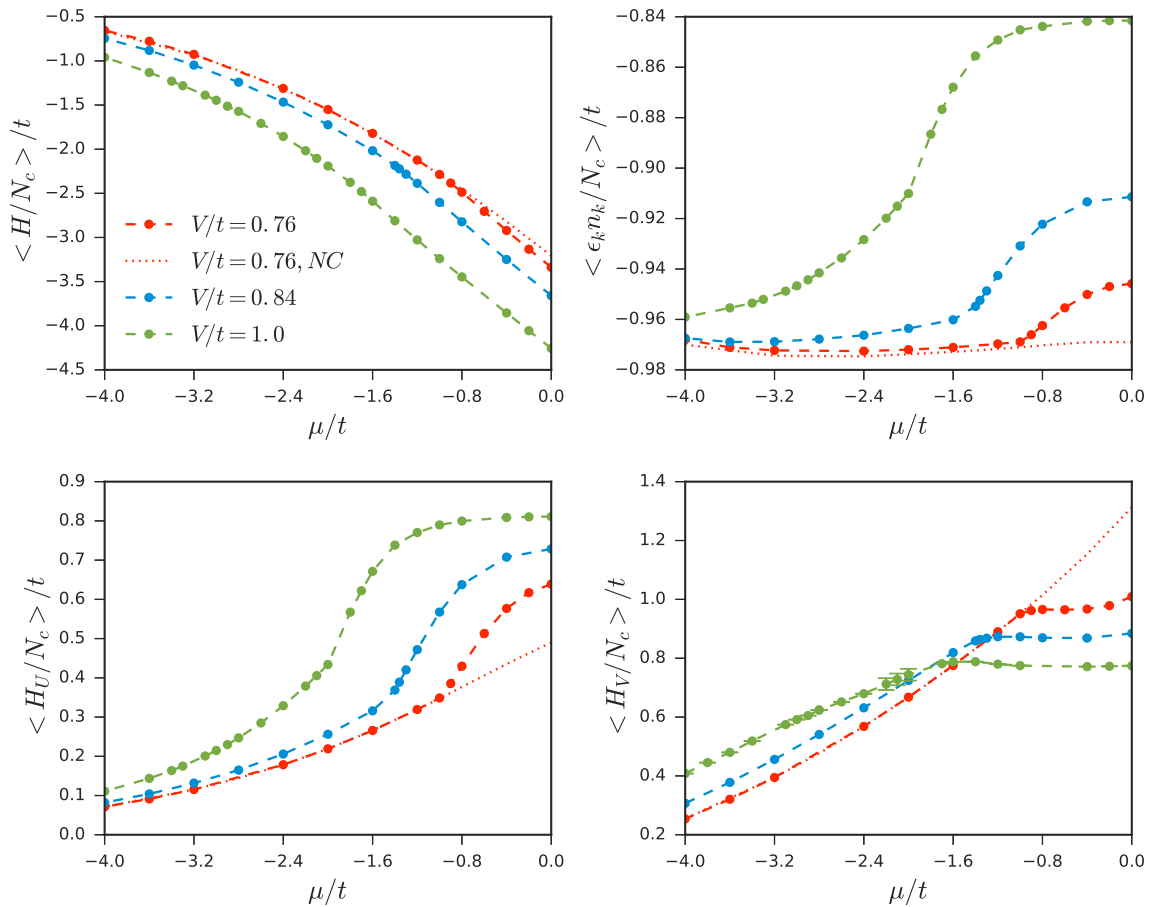}
\caption{Energetics. Top left panel: total energy per particle. Top right panel: kinetic/single particle energy. Bottom left panel: on-site contribution to the potential energy. Bottom right panel: Non-local contribution to the potential energy. Dotted red lines for $V/t=0.76$  denote the metastable (NC) solution where charge order is suppressed. Error bars, where indicated, denote errors larger than the symbol size.}
\label{fig:energy}
\end{figure*}  

\section{Energetics}\label{sec:Energy}
 Fig.~\ref{fig:energy} presents an analysis of the energetics of the charge order transition as a function of chemical potential. The top left panel shows the total energy $E_\text{tot}$ for the four non-local interaction strengths $V$ indicated and for $U/t=2.0$, $T/t=0.32$. Also shown, as dashed line, is the isotropic (NC) state where charge order has been artificially suppressed. The phase transition is visible in the top left panel as a slight change of slope and as a deviation between the symmetry broken and the isotropic state.
 
The total energy consists of two parts, a single-particle `kinetic' part $\frac{1}{N_c}\sum_K(\epsilon_K-\mu)n_K$ (note that different practitioners use different definitions of the `kinetic' energy) and an interacting part consisting of the contributions from local and non-local interaction terms. The top right panel shows the doping evolution of the kinetic part. The phase transition is clearly visible in the data, indicating that the large kinetic energy term at low doping is rapidly reduced upon entering the ordered phase. Nevertheless, the contribution of this term to the total energy is small in comparison to the interaction contribution.

The bottom two panels disentangle the interaction contributions to the total energy. The bottom left panel shows contributions from the local interaction $U$, while the bottom right panel shows contributions from the non-local interaction $V$. Note the overall magnitude of the change in comparison to the kinetic part. We first focus on the local interaction energy contribution $H_U$. The charge order insulator has a high double occupancy at half filling, and therefore a large contribution of the local interaction energy. As the charge order is melted by doping, the double occupancy is rapidly reduced and therefore the local interaction energy contribution is reduced. This behavior of the energetics is opposite from what would be expected in a Mott insulator, where the double occupancy is generally rapidly suppressed upon entering the insulating state, but qualitatively similar to what is expected at an antiferromagnetic transition.

In contrast, the non-local interaction energy $H_V$ in the charge ordered phase is strongly suppressed when compared to the uniform state. Therefore, the charge-order transition mainly reduces the non-local interaction energy at the cost of increasing the local interaction energy, while the total change to the kinetic energy is much smaller.\cite{Terletska17}

\begin{figure}[tbh]
\includegraphics[width=0.85\columnwidth]{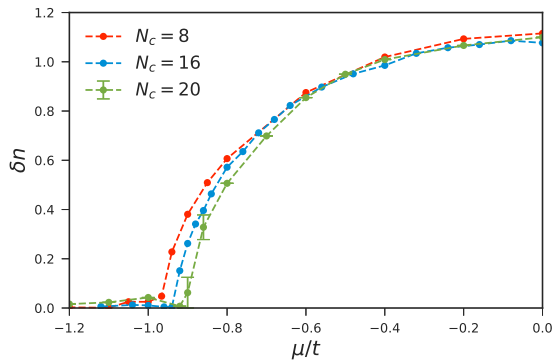}
\caption{Size of the order parameter as a function of doping as cluster size is varied from $N_c=8$ to $N_c=16$ and $N_c=20$, with $V/t=0.76$, $U/t=2$, $T/t=0.32$. For these parameters, a variation in the critical doping of around $5\%$ is visible. Away from the critical point, the order parameter quickly converges with system size.}
\label{fig:finitesize}
\end{figure}  

\section{Estimation of finite size effects}\label{sec:FS}
The dynamical cluster approximation is controlled, in the sense that $1/N_c$ is a small parameter. Away from criticality, local observables such as the order parameter or the total energy per particle are expected to converge $\sim 1/N_c.$ At criticality, where the correlation length is expected to be much larger than the system size, convergence is expected to be slow. Within our approximation, we cannot perform a rigorous finite size scaling at the critical temperature. However, from a limited range of relatively small cluster sizes we can estimate the variation of the critical region with cluster size and illustrate the variance of quantities such as the energy or the order parameter. Fig.~\ref{fig:finitesize} shows such a study for the order parameter and the critical temperature on clusters of size $8$, $16$, and $20$. Visible are deviations on the order of $5\%$ in the location of the critical doping. The size of the order parameter obtained on the larger clusters quickly converges to the value obtained for $N_c=8$. A rigorous finite size extrapolation, as it is done in the context of high-temperature cold atom calculations, is not possible for these parameters with current techniques.

\section{Conclusions}\label{sec:Conc}
In conclusion, we have performed a comprehensive study of the finite temperature phase diagram of the charge ordered phase away from half-filling of the two-dimensional extended Hubbard model using an eight-site DCA. We have studied the behavior of the order parameter $\delta n$ and constructed the phase boundaries as a function of doping $x$, local interaction $U$, nearest-neighbor interaction $V$, temperature $T$, and chemical potential $\mu$ in detail and have delineated regimes in which the system forms a half-filled charge-ordered incompressible insulator, a compressible ('metallic') charge-ordered state, and a Fermi liquid metal. We have demonstrated that charge order survives doping away from half-filling up to around $20\%$ at lower temperatures, and that increasing temperature causes the destruction of the charge order.  We have shown that the evolution of the spectral function as a function of doping shows interesting spectral weight transfer from below the gap to above the gap upon doping, and we have analyzed the changes to the various energetic contributions to the system as charge order is established or destroyed.

Our study was limited to intermediate interaction strengths ($U/t$ up to 4), and excluded infinite-range antiferromagnetic fluctuations but included short-range fluctuations of all types. Some of this limitation was due to the fact that quantum Monte Carlo impurity solvers for systems with non-local interactions suffer from a strong sign problem, even at half filling. In the future, it will be interesting to use advances in Monte Carlo methods and many-body theory\cite{Zgid17} to use this formalism to examine the effect of strong non-local interactions on the Mott transition, and to examine the putative quantum critical behavior of the extended Hubbard model at low temperature.

\acknowledgments{
This work was supported by the Simons Foundation via the Simons collaboration on the many-electron problem. JP was supported by NSF DMR-1606348. Computer time was provided by XSEDE, which is supported by National Science Foundation grant number ACI-1548562.}
\bibliographystyle{apsrev4-1}
\bibliography{EHM}

\end{document}